\documentclass[aps, prd, amsmath, floats, floatfix, twocolumn,
superscriptaddress, nofootinbib, showpacs]{revtex4}
\usepackage{graphicx}
\usepackage{epsfig}
\usepackage{color}
\usepackage{soul}
\usepackage{url}
\usepackage{bm}         % bold math symbols
\usepackage{times}

\newcommand{\beq}{\begin{equation}}
\newcommand{\eeq}{\end{equation}}
\newcommand{\beqn}{\begin{eqnarray}}
\newcommand{\eeqn}{\end{eqnarray}}

\usepackage{color}

\newcommand{\Caltech}{\affiliation{Theoretical Astrophysics 350-17,
    California Institute of Technology, Pasadena, California 91125, USA}}
\newcommand{\Cornell}{\affiliation{Center for Radiophysics and Space
    Research, Cornell University, Ithaca, New York, 14853, USA}}
\newcommand{\WSU}{\affiliation{Department of Physics \& Astronomy,
	Washington State University, Pullman, Washington 99164, USA}}
\newcommand{\CITA}{\affiliation{Canadian Institute for Theoretical 
    Astrophysics, University of Toronto, Toronto, Ontario M5S 3H8, Canada}}

\usepackage{graphicx}% Include figure files
\usepackage{dcolumn}% Align table columns on decimal point
\usepackage{bm}% bold math
\usepackage{epsf}

\begin{document}

\title{Black hole-neutron star mergers for $10M_\odot$ black holes}

\author{Francois Foucart} \Cornell \CITA%
\author{Matthew D. Duez} \WSU %
\author{Lawrence E. Kidder} \Cornell %
\author{Mark A. Scheel} \Caltech %
\author{Bela Szilagyi} \Caltech %
\author{Saul A. Teukolsky} \Cornell %

\begin{abstract}
General relativistic simulations of black hole-neutron star mergers
have currently been limited to low-mass black holes ($M_{\rm BH}
\leq 7M_\odot$), even though population synthesis models indicate
that a majority of mergers might involve more massive black holes
($M_{\rm BH} \geq 10 M_\odot$). We present the first general
relativistic simulations of black hole-neutron star mergers with
$M_{\rm BH} \sim 10M_{\odot}$. For massive black holes, the tidal
forces acting on the neutron star are usually too weak to disrupt
the star before it reaches the innermost stable circular orbit of
the black hole. Varying the spin of the black hole in the range
$a_{\rm BH}/M_{\rm BH}=0.5$ -- 0.9, we find that mergers result in the
disruption of the star and the formation of a massive accretion disk
only for large spins $a_{\rm BH}/M_{\rm BH} \geq 0.7$ -- 0.9. From
these results, we obtain updated constraints on the ability of
BHNS mergers to be the progenitors of short gamma-ray bursts as a
function of the mass and spin of the black hole. We also discuss
the dependence of the gravitational wave signal on the black hole
parameters, and provide waveforms and spectra from simulations
beginning 7 -- 8 orbits before merger.
\end{abstract}


\maketitle

\section{Introduction}
\label{intro}

The study of black hole-neutron star (BHNS) mergers offers an
opportunity to observe a general relativistic system in the strong
field regime, in the presence of material at supernuclear density,
magnetic fields, shocks, and intense neutrino radiation. As for other
compact binaries, recent interest in black hole-neutron star systems
is due in large part to their potential as sources of gravitational
waves detectable by the next generation of ground-based detectors,
Advanced LIGO~\cite{LIGO} and VIRGO~\cite{VIRGO}.  Indeed, whether
for detection or parameter estimates, these detectors require
theoretical templates in order to extract the gravitational wave
signal from the detector noise.  And in the orbits immediately
preceding the merger, as during the merger itself, these binaries
can only be properly modeled by numerical simulations in general
relativity.

The emission of gravitational waves is not the only interesting
feature of BHNS mergers, however. For compact binaries containing
at least one neutron star the outcome of the merger is by itself
an important question. The formation around the remnant black hole
of a massive, thick, and hot accretion disk from the remains of
a tidally disrupted neutron star offers ideal conditions to power
short-hard gamma ray bursts (SGRB). This possibility, combined with
the association of SGRBs with relatively old stellar populations,
makes BHNS and NS-NS mergers some of the most likely progenitors
of SGRBs (see e.g.~\cite{2007NJPh....9...17L} for a review).

Black hole-neutron star binaries have not yet been as widely
studied as NS-NS or BH-BH systems. The first simulation
of a BHNS merger in general relativity was done in 2006 by
Shibata \& Uryu~\cite{Shibata:2006ks}. Since then, various
groups have studied the influence on the dynamics of the
merger and the gravitational wave signal of the black hole
spin~\cite{Etienne:2008re,2011PhRvD..83b4005F,PhysRevD.84.064018},
the mass
ratio~\cite{Etienne:2008re,Shibata:2009cn,PhysRevD.84.064018},
the neutron star equation of
state~\cite{Shibata:2009cn,Duez:2010a,2010PhRvD..82d4049K,PhysRevD.84.064018},
the eccentricity of the orbit~\cite{2011ApJ...737L...5S}, and the
presence of a magnetic field~\cite{2010PhRvL.105k1101C}. A review
of those results can be found in~\cite{2010CQGra..27k4002D}. These
simulations indicate that mergers of low-eccentricity BHNS binaries
show two potential qualitative behaviors. In the first, the neutron
star is tidally disrupted before it reaches the innermost circular
orbit (ISCO) of the black hole. Material from the star is then
either quickly accreted onto the black hole, or ejected in a long
tidal tail. As material in the tail falls back onto the black hole,
it then forms an accretion disk. These disks are usually fairly
thick, have temperatures of a few MeV, can contain a significant
fraction of the initial mass of the star (a few tenths of a solar
mass), and maintain a baryon-free region along the rotation axis
of the black hole in which relativistic jets could be launched.
They therefore offer promising conditions for the generation of SGRBs. The
second scenario lets the star reach the ISCO before it overflows
its Roche lobe. No tidal tail is formed, and the black hole and
neutron star merge directly. Numerical simulations have confirmed
that disk formation is favored by a high black hole spin (as the ISCO
is closer to the black hole), a large neutron star (the star is more
easily disrupted), and a small black hole mass (the size of the Roche
lobe at the ISCO relative to the radius of the neutron star is 
smaller for lower mass black holes, thus making Roche-lobe 
overflow easier).

All existing simulations of BHNS mergers in general relativity
have studied black holes of mass between 2 and 5 times the mass
of the neutron star. For such configurations, disk formation
is the most likely scenario. It can only be prevented for
anti-aligned spins~\cite{Etienne:2008re,PhysRevD.84.064018},
and compact stars or massive non-spinning black
holes~\cite{2010PhRvD..82d4049K,Etienne:2008re}. However, the
choice of such low mass ratios is not very well motivated. Population
synthesis models tend in fact to favor distributions in which a large
fraction of BHNS binaries have a more massive black hole ($M_{\rm BH}\geq
10M_{\odot}$), especially for binaries formed in low-metallicity
environments~\cite{2008ApJ...682..474B,2010ApJ...715L.138B}. These
predictions come with large uncertainties, but it still appears
important to study the behavior of BHNS systems for higher mass
black holes.

Simulations of binary systems with high mass ratios are typically
more costly than their equal-mass counterparts. The reason is that
before disruption the time step of the evolution is limited by
the minimum spacing of the numerical grid (Courant-Friedrichs-Lewy
condition), which scales as the size of the smaller object. High
mass ratio simulations thus require significantly more time steps
per orbit if we want to maintain the same accuracy. Mergers
are also more difficult in this case: the rapid accretion
of matter onto the black hole from a relatively small object
can only be resolved if the numerical grid around the black
hole has an accordingly small spacing, and a small time step
is thus required as well. These increasing costs, plus the
fact that lower mass ratios offer richer physical effects, have
led to a focus on low black hole masses in general relativistic
simulations of BHNS mergers up till now. High-mass ratio BHNS binaries have,
however, been evolved using approximate treatments of gravity (e.g.
in~\cite{Lee:1998qk,Janka:1999qu,2005ApJ...634.1202R,2008ApJ...680.1326R}).
These simulations tend to indicate that high spins are required
for disk formation to be possible. However, approximate
simulations have been known
to disagree with results in full general relativity in the past, especially on
predictions regarding the mass of the accretion disk or the amount
of unbound material ejected in the tidal tail.

In this paper, we focus on the features of BHNS mergers for black
holes of mass $7-10M_{\odot}$. In Section~\ref{SpEC}, we review our
numerical code and the specific challenges related to the evolution
of high mass ratio systems, and describe recent improvements to our
code. We also present updated diagnostics regarding the accuracy of
our gravitational waveforms, and the influence of gauge choices on
our results. Section~\ref{sec:ID} describes the initial conditions
for our simulations, while Section~\ref{sec:NR} presents our
numerical results. In particular, we show that for massive black
holes ($M_{\rm BH}\geq 10 M_\odot$), disks cannot be formed unless
the spin of the black hole is high ($a_{\rm BH}/M_{\rm BH} \geq
0.7$). We also discuss the imprint of the binary parameters on the
gravitational wave signal. Finally, Section~\ref{sec:discussion}
focuses on consequences for the observation of SGRBs.

\section{Methods and Diagnostics}
\label{SpEC}

The black hole-neutron star binaries presented in this
paper are evolved using the numerical code developed by the
SpEC collaboration~\cite{SpEC}.  The coupled system formed by
Einstein's equations and the general relativistic hydrodynamics
equations is solved using the two-grid method described in Duez
et al.~\cite{Duez:2008rb}, with the improvements discussed in
Foucart et al.~\cite{2011PhRvD..83b4005F}. We refer the reader
to~\cite{Duez:2008rb} for
a detailed description of the numerical methods used in SpEC,
and only discuss here some recent modifications
improving the general performance of the code, and facilitating
the merger of high mass ratio BHNS binaries.

In the two-grid method, Einstein's equations are solved within the
generalized harmonic formalism~\cite{Lindblom:2006}
using pseudospectral methods, while the fluid equations are evolved
on a separate finite difference grid.
This allows us to take advantage of the efficiency of pseudospectral
methods for the evolution of Einstein's equations
in regions in which the solution is smooth (i.e., away from the
neutron star), while limiting the extent of the finite difference
grid to regions in which matter is present. The price to pay is that,
as the two sets of equations are coupled, source terms have to be
interpolated between the two grids at each time step.

The relativistic hydrodynamics equations are solved in conservative
form: the evolved conservative variables ${\bm U}$ satisfy
equations of
the form
\beq
\partial_t {\bm U} + \nabla \cdot{\bm F}({\bm U}) = {\bm S}({\bm U}),
\eeq
where the fluxes ${\bm F}$ and source terms ${\bm S}$ are functions
of the variables
${\bm U}$ but not of their derivatives.  A conservative shock-capturing
scheme is characterized by a reconstruction method, whereby fluid quantities
on either side of cell faces are constructed, and a Riemann solver that supplies
the resulting fluxes across each cell face.  For these simulations, we use
5th-order WENO reconstruction~\cite{Liu1994200,Jiang1996202} with the
smoothness indicator proposed by Borges~{\it et al}~\cite{Borges} and HLLE
fluxes~\cite{HLL}.

Since the publication of Foucart et al.~\cite{2011PhRvD..83b4005F},
other modifications of the SpEC code have significantly improved
the efficiency of our simulations. In the following sections,
we discuss the most important of these recent changes.

\subsection{Adaptive time stepper}

In previously published simulations, black hole-neutron star binaries
were evolved using a fixed time step $\Delta t=C \Delta x_{\rm
min}$, where $\Delta x_{\rm min}$ is the smallest grid spacing
and $C$ some constant chosen so that the evolution satisfies the
Courant-Friedrichs-Lewy (CFL) stability condition at all times. As
the truncation error in these simulations is usually dominated by
the effects of the spatial discretization, this is sufficient for
the effects of the discretization in time to be negligible. However,
the resulting time step can be significantly smaller than necessary
for a large fraction of the evolution. Indeed, the maximum value of
$C$ that gives stable evolutions can change over time\footnote{The
variation in time of the CFL stability condition occurs mainly because
the pseudospectral grid follows the evolution of
the binary and, in particular, contracts as the binary spirals
in.} and is a priori unknown for the evolution of Einstein's
equations on a pseudospectral grid. Additionally, the only way to
confirm that the error in a given simulation is indeed dominated
by spatial truncation error is to rerun it with a different choice
for $C$. Using an adaptive time stepper, we instead choose at
any time the largest time step for which the time stepping
error is smaller than some tolerance.
If the CFL instability begins to grow, this shows up as an increasing
time stepping error and the time stepper automatically reduces the time
step. In practice, time step control is
done by evaluating the time stepping error through comparisons of the
result of the time evolution with what would be obtained if a lower
order scheme were used. We evolve using the 3rd-order Runge-Kutta
algorithm, and compare with the results of a second-order method. The
time step $h_{n+1}$ is then determined using a proportional-integral (PI) stepsize
control~\cite{numrec_PI,GustaffsonPI1991,HairerPI1996,SoderlindPI2003}:
\beq
h_{n+1}=Sh_n \epsilon_n^{-\alpha} \epsilon_{n-1}^\beta
\eeq 
where $\epsilon_n$ is the error measured at time step $n$, $S=0.9$
is a safety factor, $\beta\approx0.13$ and $\alpha\approx 0.23$.

For the evolution of a BHNS binary over 7 -- 8 orbits, choosing the
optimal $\Delta t$ at all times can reduce the simulation time by a
factor of $\sim 1.5-2$. Additionally, as the minimum value of $C$
giving stable evolutions is unknown, our choice when using fixed
time steps was usually overly cautious. The actual gain in the time
required to evolve BHNS systems with SpEC at a given resolution is
thus closer to a factor of 2 -- 3.

\subsection{Control system}

The evolution of a black hole on a pseudospectral grid requires the
excision of the region surrounding the singularity. As a consequence,
the numerical domain has an inner boundary $\mathcal{S}_{\rm in}$,
typically a sphere in the coordinates of the numerical grid. If
any information can enter the domain through $\mathcal{S}_{\rm in}$,
an unknown boundary condition has to be imposed there. Inside the
apparent horizon of the black hole, this can however be avoided
for `good' choices of $\mathcal{S}_{\rm in}$. More specifically,
we compute the local characteristic speeds of the generalized
harmonic equations (which are hyperbolic), and require that on
$\mathcal{S}_{\rm in}$ all characteristic speeds point out
of the numerical grid. We also require that $\mathcal{S}_{\rm in}$
remain inside the apparent horizon of the black hole. In practice,
this can be difficult when the black hole is rapidly evolving through
tidal distortion or matter accretion. During merger, we try to keep
the apparent horizon nearly spherical in the coordinates of the
numerical grid, and slightly outside $\mathcal{S}_{\rm in}$,
through the use of a coordinate map between the grid coordinates
and the inertial frame:
\beq
\tilde{r} = r (1 + \sum_{lm} Y_{lm}(\theta,\phi) c_{lm}),
\eeq
where $\tilde{r}$ is the distance from the black hole center in
the inertial frame, $r$ the same distance in the grid coordinates,
($\theta$,$\phi$) the usual angles in spherical coordinates (in both
frames), and $c_{lm}$ arbitrary coefficients used to control the
location of $\mathcal{S}_{\rm in}$ in the grid frame.
The quantity $c_{00}$
thus controls the size of $\mathcal{S}_{\rm in}$, while the other
coefficients control its shape. In previous simulations, the control
system attempted to choose the $c_{lm}$ so that the apparent horizon
remained a sphere of fixed radius in the grid frame. This proved
inadequate for high mass ratio simulations: there is a priori no way
to know at what distance the apparent horizon should be kept from
$\mathcal{S}_{\rm in}$, and making the wrong choice can easily lead
to characteristic speeds entering the domain on $\mathcal{S}_{\rm
in}$. Instead, we now choose $c_{00}$ by using the characteristic
speeds themselves as input for the control system. Indeed, $\partial_t
c_{00}$
affects the velocity of the surface $\mathcal{S}_{\rm in}$ in
inertial  coordinates, and can thus be used to increase or decrease
the characteristic speeds at will. The other coefficients are still
chosen so that the apparent horizon remains spherical.

As the time scale over which $\mathcal{S}_{\rm in}$ needs to change
varies by orders of magnitude between the early orbits and the merger,
the control system used to fix the $c_{lm}$ has also been modified in order
to choose adaptively the time scale over which it attempts to damp
any deviation of the motion of $\mathcal{S}_{\rm in}$ from its
desired behavior. 
%Descriptions of the methods used by our control
%system to fix its damping time scales, as well as more details on
%the control of $\mathcal{S}_{\rm in}$ through characteristic speeds
%can be found in[].
%\textcolor{red}{Ref? Mark: Do you want to write more /
%less on the control system here?}

\subsection{Adaptive Mesh Refinement on the spectral grid}

Another complication arising for high mass ratio mergers is the
rapid motion of the high-density neutron star core across the
numerical grid as it falls into the black hole. This high-density
material generates large source terms in Einstein's equations,
which vary quickly and can have very steep gradients. Maintaining
a grid resolution high enough to resolve these features in all
parts of the numerical domain where they might occur, and during
the entire simulation, is far too costly to be realistic. In
practice, we need to adapt the numerical domain at regular time
intervals in order to increase resolution in regions where the
neutron star core is located. This was already done on the finite
difference grid, by limiting the grid to regions in which matter
is present~\cite{2011PhRvD..83b4005F}. The possibility of modifying
the pseudospectral grid during a simulation, on the other hand, is
a new feature of our code which was initially developed to improve
the accuracy of black hole binary mergers.
%\textcolor{red}{\bf Add Ref. here!}
The choice of how to modify the resolution in any given subdomain of
the numerical grid relies on estimates of the truncation error
based on the fall-off rate of the coefficients of the spectral
expansion. For a smooth solution, these coefficients should fall
off exponentially. In BHNS simulations, the solution is not smooth,
but the metric quantities are still
continuous, and their spectral expansion converges as a power
law. The determination of the required resolution is less reliable
than in vacuum---and better methods to control that choice should
be found if we want to take full advantage of the method---but
in its current form this spectral adaptive mesh refinement
(AMR) technique has already allowed us to evolve configurations
that were otherwise too costly to simulate accurately. In this
paper, spectral AMR is used for the merger of case Q7S5, the
only configuration in which the neutron star merges without any
significant tidal disruption.
%Fig.~\ref{fig:AMR} shows the numerical grid and the density profile during that merger. As spectral AMR currently does not change the distribution of subdomains on the grid, the area around the initial location of the neutron star remains over resolved --- but otherwise, the grid spacing has been increased on the side of the black hole in which large gradients in the stress-energy tensor exist, and remains larger elsewhere. 
%\textcolor{red}{Bela: Anything to change/add here?}

%\begin{figure}
%\caption{Spectral grid in the equatorial plane when using adaptive mesh refinement (case Q7S5).}
%\label{fig:AMR}
%\includegraphics[width=8.5cm]{Images/MergerAMR1}
%\end{figure}

\subsection{Gravitational Wave Accuracy}

Gravitational wave templates for ground based detectors such
as Advanced LIGO require accurate waveforms. Even the longest
and most accurate simulations of binary black holes combined
with Post-Newtonian results to generate hybrid waveforms might
currently be insufficient for precise parameter estimates
(i.e., the accuracy of the parameters might be limited by
the quality of the templates instead of the noise in the LIGO
data)~\cite{PhysRevD.84.064013,2011CQGra..28m4002M}.
Templates could well
require a phase accuracy in the numerical waveform of $0.1\,{\rm
rad}$ over significantly more than the 30 cycles of the longest
available numerical waveform~\cite{2011CQGra..28m4002M}. Accuracy
requirements are lower for detection purposes, and some
parameters could be usefully constrained with waveforms at the
current level of accuracy (e.g., the neutron star radius from NS-NS
mergers~\cite{2009PhRvD..79l4033R}). But it is clear that template
accuracy is a significant concern for gravitational wave astronomy.

Binary neutron stars and BHNS binaries are particularly
challenging, as the need to evolve the neutron star fluid
adds new sources of errors to the simulations. In previous
papers~\cite{Duez:2010a,2011PhRvD..83b4005F}, we evolved
the system for only 2 -- 3 orbits before merger, and the phase error in
the gravitational wave signal was of the order of a few tenths
of a radian even before merger. The simulations presented here
are significantly longer (7 -- 8 orbits), and more accurate (the
'low' resolution used here is roughly equivalent to our standard
resolution in~\cite{2011PhRvD..83b4005F}). We thus need to reassess
the accuracy of our waveforms.

\begin{figure}
\caption{Phase error $\delta \phi$ for the (2,2) mode of the gravitational
wave signal as a function of the number of orbits, for simulation
Q7S7. The wave is extracted at finite radius $R=100M_{\rm total}$,
and we do not attempt to shift it in phase or time.}
\label{fig:dphi}
\includegraphics[width=8.3cm]{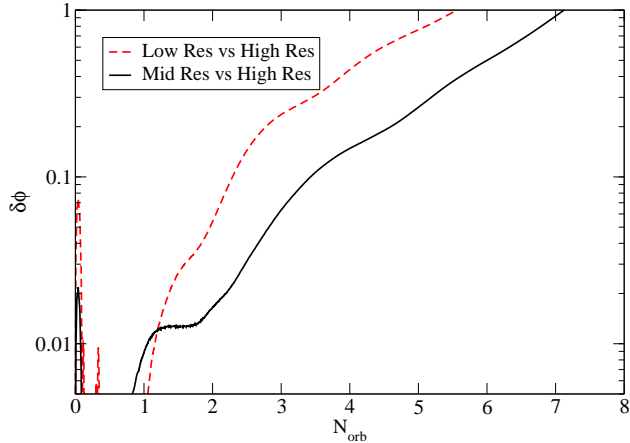}
\end{figure}

The error in the gravitational wave signal is evaluated using the
Q7S7 simulation ($q=7$, $a_{\rm BH}/M_{\rm BH}=0.7$), which we run at
3 different resolutions. The gravitational wave signal is extracted
at finite radius $R=100 M_{\rm tot}$, where $M_{\rm tot}=M_{\rm
BH}+M_{\rm NS}$. We directly compute the strain $h$ expended
in spherical harmonics ($h=\sum_{lm} h_{lm} Y_{lm}$), using the
Regge-Wheeler-Zerilli method \cite{1957PhRv..108.1063R,1970PhRvL..24..737Z,
2001PhRvD..64h4016S,2009CQGra..26g5009R}.
In Fig.~\ref{fig:dphi}, we
show the raw phase difference between our high resolution simulation
and the lower resolution simulations for the dominant $h_{22}$
mode. The phase is directly extracted from the real and imaginary
parts of $h_{22}$, without any time or phase shift. The phase error
in the low resolution case reaches $0.2\,{\rm rad}$ after only 2.5 orbits,
as expected from the fact that it is equivalent to the resolution
used in~\cite{2011PhRvD..83b4005F}. At higher resolution, however,
the same phase difference is only reached after approximately $4.5$ orbits,
and the overall accuracy is significantly better up to merger
even though by then the phase shift is of the order of a radian.

\begin{figure}
\caption{(2,2) mode of the strain {\it h} for simulation
Q7S7 at our two highest resolutions, extracted at finite radius
$R=100M_{\rm total}$. We shift the signal in time and phase so
that the phase at the peak of the signal is $\phi=0$, and the time
$t=0$. }
\label{fig:dphishift}
\includegraphics[width=8.3cm]{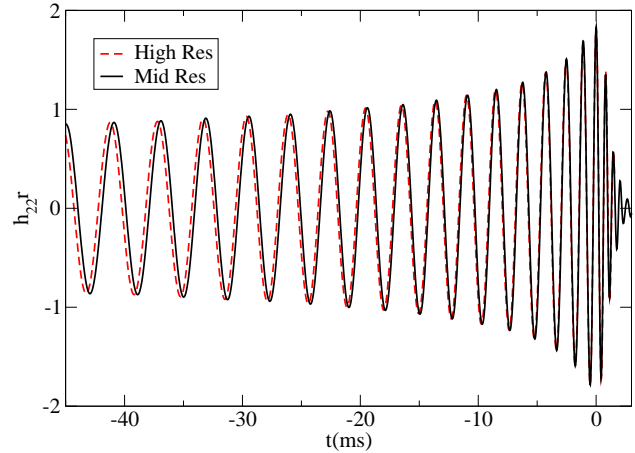}
\end{figure}

Such errors are typical of current general relativistic
simulations. We can, for example, compare our results
with those of the most recent BHNS mergers of Kyutoku et
al.~\cite{PhysRevD.84.064018}. There, the error in the waveform is
measured after applying a shift in time and phase. This is justified
from an observational point of view by the fact that a detector
would not be sensitive to such shifts, even though it might
also give an optimistic view of the uncertainties at the time of
merger. Fig.~\ref{fig:dphishift} shows the wave from the two highest
resolution of simulation Q7S7, but shifted so that the time at which
the signal peaks is $t=0$ and the phase at that time is $\phi=0$. The
errors appear comparable to Fig.25 of~\cite{PhysRevD.84.064018},
and significantly smaller than the differences due to changes in
the parameters of the binary, discussed in Sec.~\ref{sec:GW}.

\subsection{Gauge dependence}

In the generalized harmonic formulation, the gauge is specified through a
choice of the variable $H_a=g_{ab}\nabla_c\nabla^cx^b$.  For most runs
reported here, we evolve this function as in our previous
work~\cite{2011PhRvD..83b4005F}. 
During the inspiral, we fix $H_a$ in the comoving frame at its
initial value.  During the merger, we continue
to fix $H_a$ near the black hole; away from the hole, we damp it to zero. 
Because this gauge is presumably not optimal for the merger evolution
near the black hole, significant grid stretching is expected and found
during this part of the evolution.  To test whether the gauge behavior
significantly affects the accuracy of the simulations, we reran the
merger phase of the Q5S5 runs using the damped-wave gauge introduced
by Szil{\'a}gyi, Lindblom, and Scheel~\cite{Szilagyi:2009qz}.  The
resulting metric evolution is, of course, different (and noticeably
more moderate), but the gravitational wave signal should be the same. 
To within numerical errors, this is indeed what we find.  Also, the
difference in waveform between resolutions is comparable for each gauge,
indicating that grid distortion is not the major source of waveform error.

\section{Initial configurations}
\label{sec:ID}

We generate initial data for our simulations with the elliptic
solver SPELLS~\cite{Pfeiffer:2003a}. To obtain realistic initial
configurations for the evolution of black hole-neutron star binaries,
we have to satisfy the constraints in Einstein's equations, as
well as choose a realistic initial state for the fluid forming the
neutron star. For the fluid we require hydrostatic equilibrium
as well as an irrotational velocity profile. The procedure for solving
for these conditions and obtaining compact objects with the desired
masses, spins and orbital parameters is detailed in Foucart et
al.~\cite{Foucart:2008a}.
The constraint equations are solved in the extended conformal thin
sandwich formalism~\cite{York:1999,Pfeiffer:2003}. In that formalism,
the metric is decomposed as
\beq
ds^2 = g_{\mu \nu} dx^\mu dx^\nu = -\alpha^2 dt^2 + \phi^4 \tilde\gamma_{ij}
(\beta^i dt + dx^i)(\beta^j dt + dx^j)
\eeq
where $\alpha$ is the lapse function, $\beta^i$ the shift
vector, $\tilde \gamma_{ij}$ the conformal 3-metric and $\phi$
the conformal factor. The constraints can then be expressed as
elliptic equations for $(\alpha\phi, \beta^i, \phi)$. The choice of
the conformal metric, the trace of the extrinsic curvature $K_{\mu
\nu}=-\frac{1}{2} \mathcal{L}g_{\mu \nu}$, and their time derivatives
is then arbitrary. Together with the boundary conditions and the
stress-energy tensor, these arbitrary choices
will determine the characteristics of the
spacetime under consideration. A standard choice in the literature
is $\tilde g_{ij} = \delta_{ij}$, $K=0$. However, for high black
hole spins this leads to configurations far from equilibrium,
and potentially large modifications of the black hole spin and
mass at the beginning of the simulation. We prefer to choose the
free data $(g_{ij},K)$ according to the prescription of Lovelace
et al.~\cite{Lovelace:2008a}: close to the horizon of the BH, the
metric matches the Kerr-Schild solution. (For the adaptation of that
method to BHNS systems, see~\cite{Foucart:2008a}.)

In this paper, we are interested in the behavior of BHNS binaries
for high mass ratio $q=M_{\rm BH}/M_{\rm NS} =7$. We consider 4
different configurations, summarized in Table~\ref{tab:ID}. The
first is at a slightly lower mass ratio $q=5$, a range of masses
already explored by Etienne et al.~\cite{Etienne:2008re},
Kyutoku et al.~\cite{PhysRevD.84.064018} and Chawla et
al.~\cite{2010PhRvL.105k1101C}. The spin of the black hole is $a_{\rm
BH}/M_{\rm BH}=0.5$, aligned with the orbital angular momentum
of the binary. The other three simulations use $q=7$. We show in
Sec.~\ref{sec:mergers} that these cases span the range of BH spins over
which the qualitative features of the merger vary: $a_{\rm BH}/M_{\rm
BH}=0.5,0.7,0.9$. All configurations are started approximately $ 8$ orbits
before disruption, and have low eccentricity $e \leq 0.005$ (obtained
using the iterative method developed by Pfeiffer et al.~\cite{Pfeiffer:2007a}). For
the equation of state of nuclear matter, we choose a polytrope:
the pressure $P$ and internal energy $\epsilon$ are expressed as
functions of the baryon density $\rho_0$ and temperature $T$:
\beqn
P&=&\kappa \rho_0^\Gamma +\rho_0 T\\
\epsilon &=& \frac{P}{\rho_0(\Gamma-1)},
\eeqn
with $\Gamma=2$ and $\kappa$ chosen so that the compaction of
the star is $C=M/R=0.144$. This corresponds to a stellar
radius $R\sim 14.4\,{\rm km}$ for a $1.4M_\odot$ neutron star. Recent
predictions regarding the radius of neutron stars indicate that
this is probably close to an upper bound on the radius of real stars
(see, e.g.,~\cite{2010PhRvL.105p1102H}). 

\begin{table}
\caption{Initial configurations studied. $a_{\rm BH}/M_{\rm BH}$
is the dimensionless BH spin, $M_{\rm BH,NS}$ the ADM masses of equivalent isolated BH
and NS, $R_{\rm NS}$ the radius of the star, $\Omega_{\rm rot}^0$
the initial angular frequency of the orbit, $M_{\rm tot}$ the ADM
mass of the system at infinite separation, and $e$ the initial
eccentricity. The number of orbits $N_{\rm orbits}$ is measured
at the point at which $10\%$ of the mass has been accreted by the
black hole.}
\label{tab:ID}
\begin{tabular}{|c|c|c|c|c|c|c|}
\hline
 Name & $\frac{M_{\rm BH}}{M_{\rm NS}}$ & $\frac{a_{\rm BH}}{M_{\rm BH}}$ &
$C=\frac{M_{\rm NS}}{R_{\rm NS}}$ & $\Omega_{\rm rot}^0 M_{\rm tot} $ &
$N_{\rm orbits}$ & e\\
 \hline
Q7S5 & 7 & 0.5 & 0.144 & 3.84e-2 & 7.7 & 0.001\\
 \hline
Q7S7 & 7 & 0.7 & 0.144 & 4.14e-2 & 8.3 & 0.004\\
 \hline
Q7S9 & 7 & 0.9 & 0.144 & 4.47e-2 & 8.8 & 0.005\\
%\hline
%Q7S9c & 7 & 0.9 & 0.175 & 4.48e-2 & \\
 \hline
Q5S5 & 5 & 0.5 & 0.144 & 3.55e-2 & 7.2 & 0.003\\
 \hline
\end{tabular}
\end{table}

We use three different resolutions: $80^3$, $100^3$ and $120^3$
points on the finite difference grid (which only covers the neutron
star), and
$56.9^3$, $64.5^3$ and $72.0^3$ points on the pseudospectral
grid. At the highest resolution on a moderate size cluster, 8 orbits
require approximately $40,000$ CPU-hrs
(about a month on 48 processors). During mergers, the resolution
of both the finite difference and pseudospectral grids is higher,
and depends on the configuration studied.

\section{Results}
\label{sec:NR}

\subsection{Mergers}
\label{sec:mergers}

\begin{figure*}
\caption{Matter distribution for simulation Q7S5. {\it
Left:} Most of the neutron star material plunges directly
into the black hole at the time of merger. Density contours:
$\rho_0=(10^{14},10^{13},10^{12},10^{11},10^{10})\, {\rm g/cm^3}$. {\it
Right:} 1ms later, only $2\%$ of the neutron star material
remains outside the black hole, and most of it
will rapidly accrete onto the hole. Density contours:
$\rho_0=(10^{12},10^{11},10^{10})\, {\rm g/cm^3}$.}
\includegraphics[width=8cm]{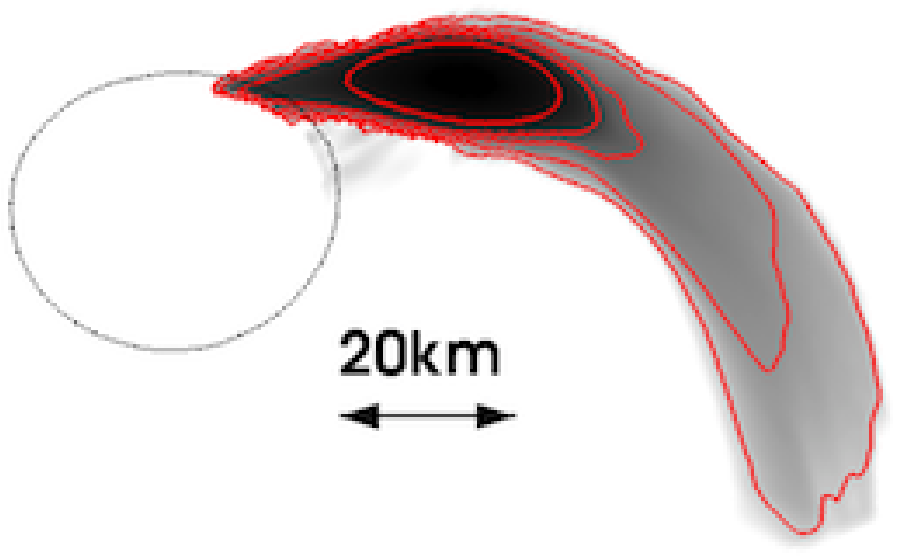}
\includegraphics[width=8cm]{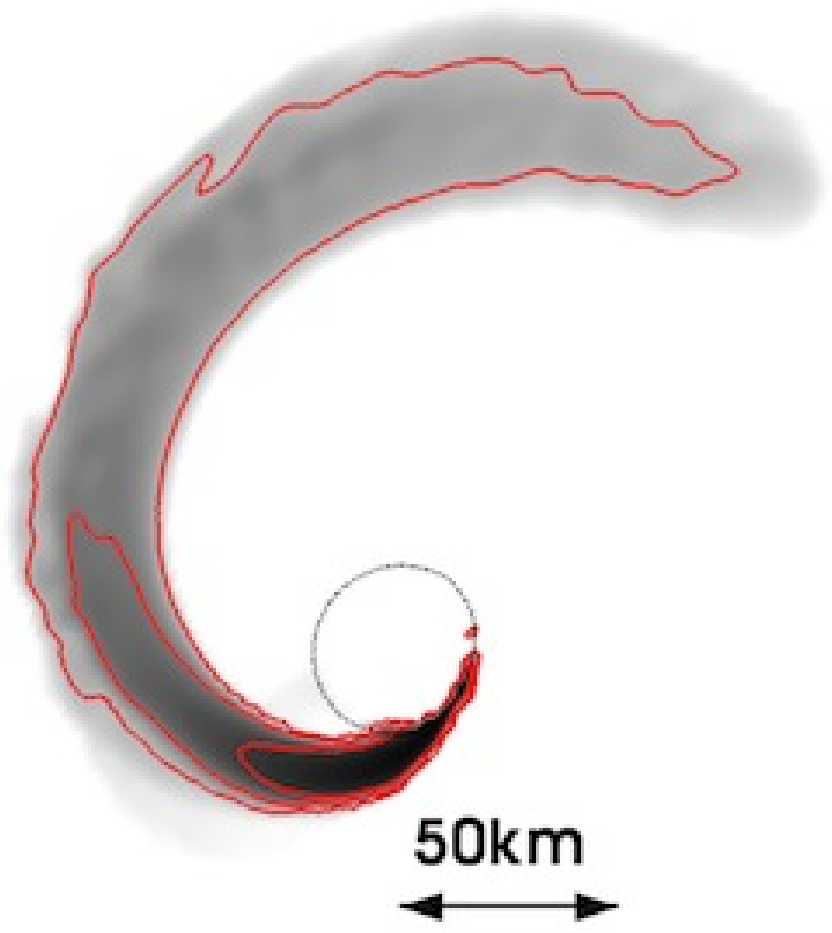}
\label{fig:Q7S5}
\end{figure*}

Most existing simulations of BHNS mergers with lower mass black
holes ($M_{\rm BH} \sim 3${} -- $7 M_\odot$) show very similar qualitative
behavior. As the neutron star gets close to the black hole,
tidal forces cause the star to overflow its Roche lobe. Most
of the matter is either rapidly accreted onto the black hole,
or ejected into an extended tidal tail. Material in the tidal
tail then rapidly forms a thick and hot accretion disk around
the black hole ($T\sim 1$ -- $5\,{\rm MeV}$). To
obtain different behaviors, one must consider either highly
eccentric orbits~\cite{2011ApJ...737L...5S}---in which case partial
disruption of the neutron star is possible---or anti-aligned black
hole spins~\cite{Etienne:2008re}, where it becomes possible to
directly accrete the entire neutron star since it reaches the innermost
stable circular orbit before overflowing its Roche lobe. Negligible
disk masses are also observed for nonspinning black holes and
fairly compact stars at the high end of the range of studied BH mass
($q=4$ -- 5)~\cite{Etienne:2008re,PhysRevD.84.064018}.

The situation changes significantly when considering more massive
black holes, $q\sim 7$. For such binaries, the standard behavior
appears to be the direct merger of the two compact objects, without
significant tidal disruption of the stars. The simulations presented
in this paper use neutron stars with radii on the high end of
the theoretically acceptable values. This is known to favor tidal
disruption~\cite{Shibata:2009cn,Duez:2010a,PhysRevD.84.064018},
as material on the surface is less tightly bound to the neutron
star. Even so, in our lower spin simulation ($q=7$, $a_{\rm
BH}/M_{\rm BH}=0.5$), $99\%$ of the material falls into the black
hole in less than $2\,{\rm ms}$. Snapshots of the matter configuration
for that simulation are given in Fig.~\ref{fig:Q7S5}. Once accretion
begins, most of the neutron star material remains within a coordinate
distance of less than about $40\,{\rm km}$ from the center of the black
hole, and only a negligible amount of matter forms a tidal tail
(approximately $0.5\%$ of the NS mass). As spin misalignment, a lower spin
and a more compact neutron star all work against tidal disruption,
this gives us a strict lower bound on the black hole spin required
to disrupt the star for black holes with mass $M_{\rm BH} \geq
7M_{\rm NS}$.

\begin{figure*}
\caption{Matter distribution for simulation Q7S7. {\it
Left:} Disruption of the neutron star. Material that is
not immediately accreted onto the black hole (about $8\%$
of the NS mass) forms a thin tidal tail. Density contours:
$\rho_0=(10^{14},10^{13},10^{12},10^{11},10^{10})\, {\rm g/cm^3}$. {\it
Center:} Disk formation, about 8ms later.  A low density accretion
disk begins to form about $50\,{\rm km}$ away from the black hole, but
rapid mass accretion from the tidal tail causes the disk profile to
vary significantly over time. The maximum density is $3\times
10^{11}\,{\rm g/cm^3}$. Density contours: $\rho_0=(10^{11},10^{10})\,
{\rm g/cm^3}$. {\it Right:} About 20ms after disruption, a stable
accretion disk has formed with low maximum density $\rho_{\rm max}
\sim 2 \times 10^{10}\,{\rm g/cm^3}$. About $2.5\%$ of the initial mass of
the neutron star remains in the disk at that time, but accretion
onto the black hole would destroy the disk within approximately $ 20\, {\rm
ms}$. Density contour: $\rho_0=10^9\,{\rm g/cm^3}$.}
\label{fig:Q7S7}
\includegraphics[width=5.7cm]{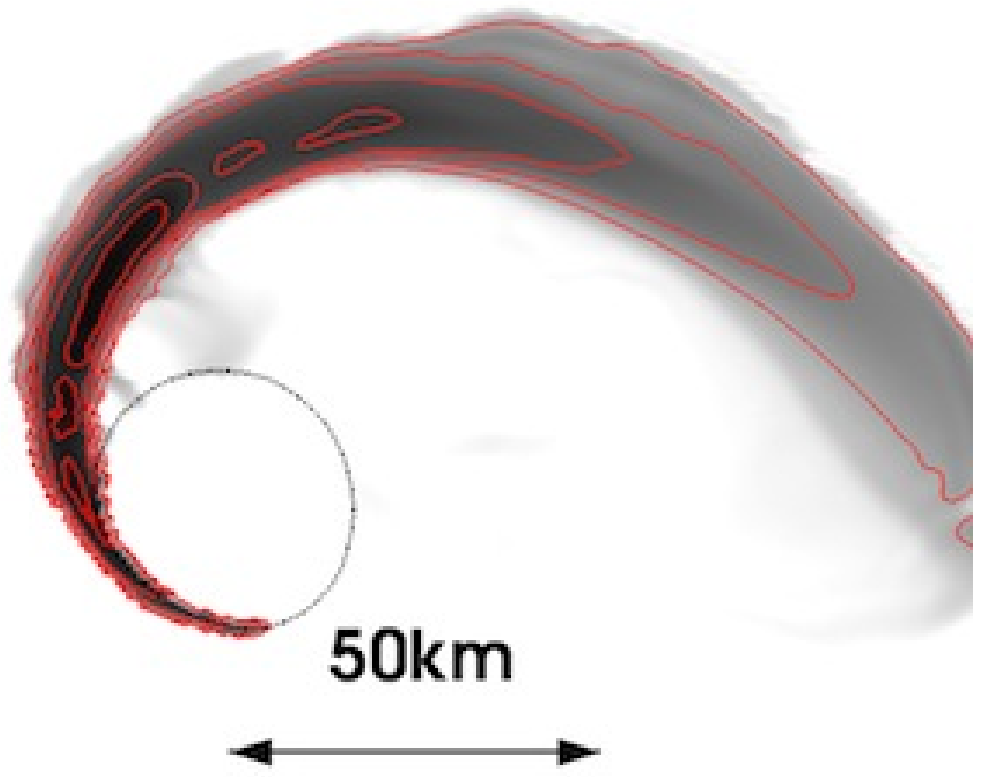}
\includegraphics[width=5.7cm]{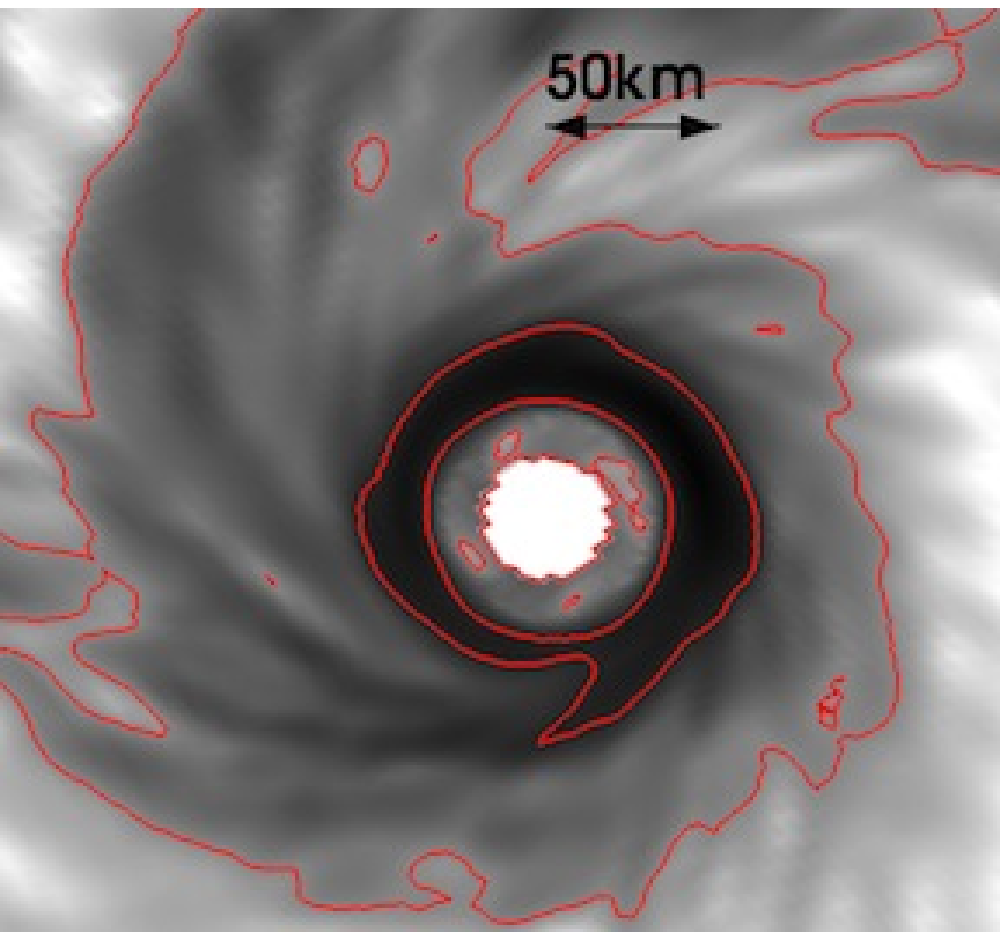}
\includegraphics[width=5.7cm]{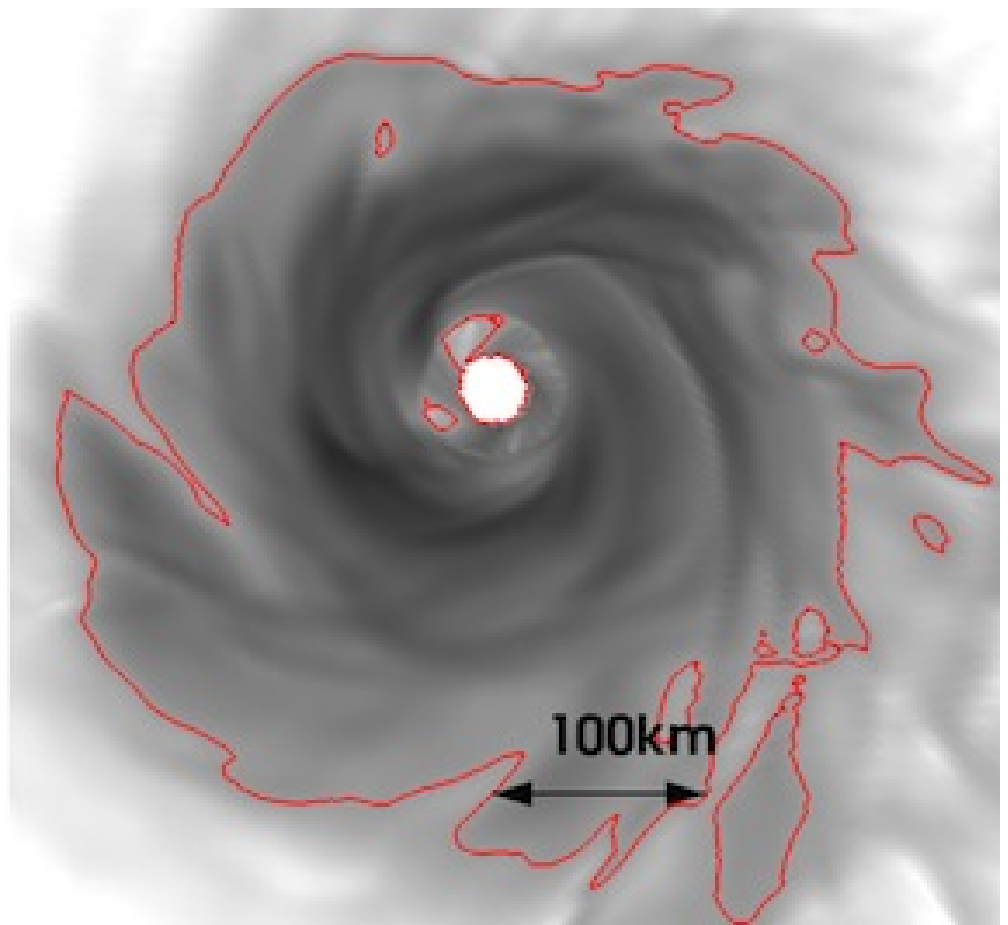}
\end{figure*}

When the spin of the black hole is increased to $a_{\rm BH}/M_{\rm
BH}=0.7$, the disruption of the neutron star becomes clearly visible
(Fig.~\ref{fig:Q7S7}, left). A long tidal tail extending more
than $100\,{\rm km}$ away from the black hole and containing about
$8\%$ of the initial mass of the star is created from the ejected
material. About $5\,{\rm ms}$ after disruption, a proto-accretion disk
forms (Fig.~\ref{fig:Q7S7}, center). However, the disk contains
only a fairly small portion of the ejected material, and is still
thinner ($H/R \sim 0.1$, where H is the half-thickness of the disk) 
and colder ($T<1\,{\rm MeV}$) than what is
usually observed in mergers with less massive black holes.

Rapid accretion from the tidal tail prevents the disk from reaching
an equilibrium profile for about $20\,{\rm ms}$. By that point, only
$2.5\%$ of the baryonic mass remains outside the black hole
(Fig.~\ref{fig:Q7S7}, right). As the radius of the disk is about 
$100\,{\rm km}$, its density is much lower than in the case of
higher black hole spins (see below) or lower black hole masses:
the maximum density is only about $2\times 10^{10}\,{\rm g/cm^3}$. The
thickness ($H/R \sim 0.3$) and temperature ($T\sim 2\,{\rm MeV}$)
of the disk are closer to the equilibrium values found for
$q=3$~\cite{2011PhRvD..83b4005F}, but this still appears to be
a less desirable configuration for the potential generation of a
short gamma-ray burst.

\begin{figure*}
\caption{Matter distribution for simulation Q7S9. {\it
Left}: Disruption of the neutron star, very similar
to the  Q7S7 case (Fig.~\ref{fig:Q7S7}), but with more
material ejected into the tidal tail. Density contours:
$\rho_0=(10^{14},10^{13},10^{12},10^{11},10^{10})\,{\rm g/cm^3}$. {\it
Center:} Disk formation 6.5ms later. More than $25\%$ of the fluid
material remains outside the black hole. A massive accretion disk
rapidly forms, containing more than $10\%$ of the matter. Density
contours: $\rho_0=(10^{12},10^{11},10^{10})\,{\rm g/cm^3}$. {\it Right:}
20ms after disruption, a massive disk remains. It is 10 times
denser than in the lower spin case, and has been in a nearly
stable configuration for more than $10\, {\rm ms}$. Density contours:
$\rho_0=(10^{11},10^{10})\,{\rm g/cm^3}$. }
\label{fig:Q7S9}
\includegraphics[width=5.7cm]{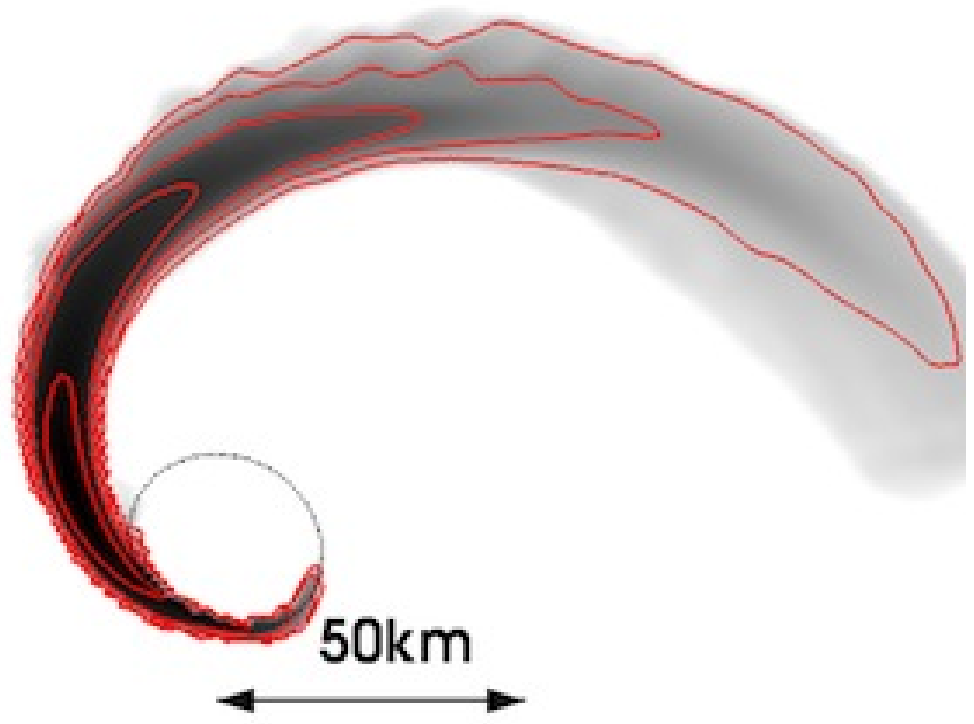}
\includegraphics[width=5.7cm]{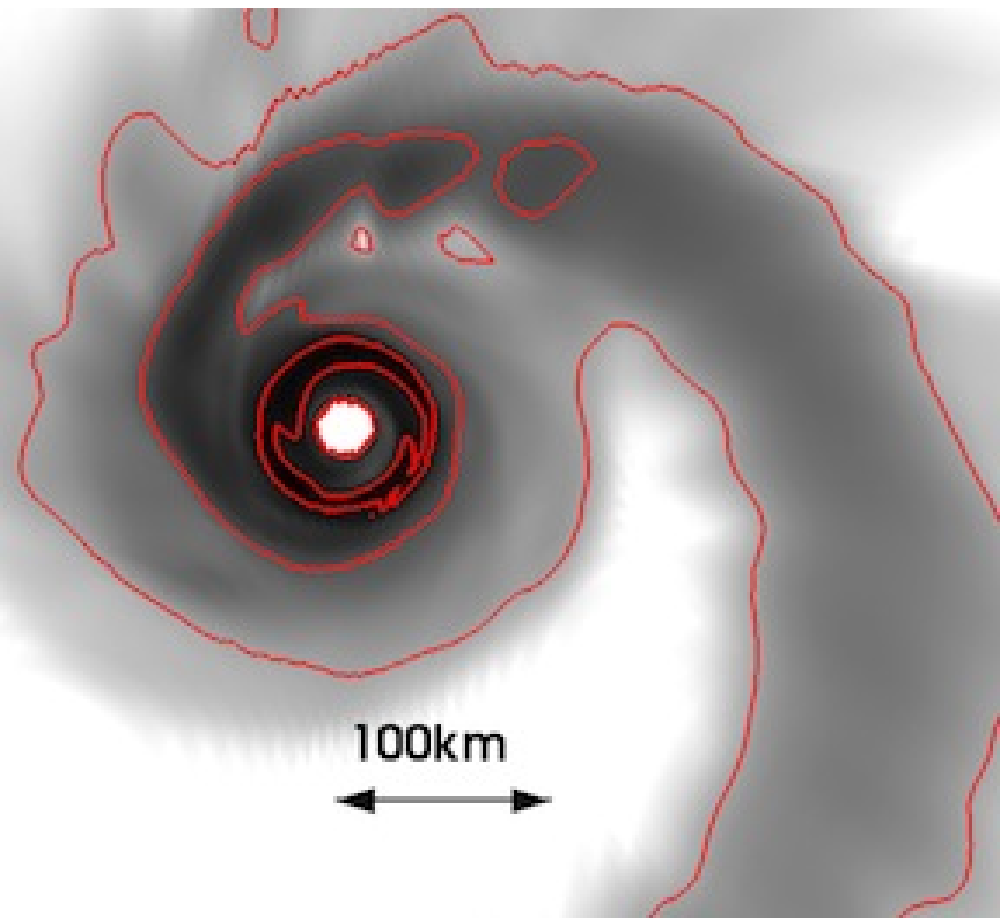}
\includegraphics[width=5.7cm]{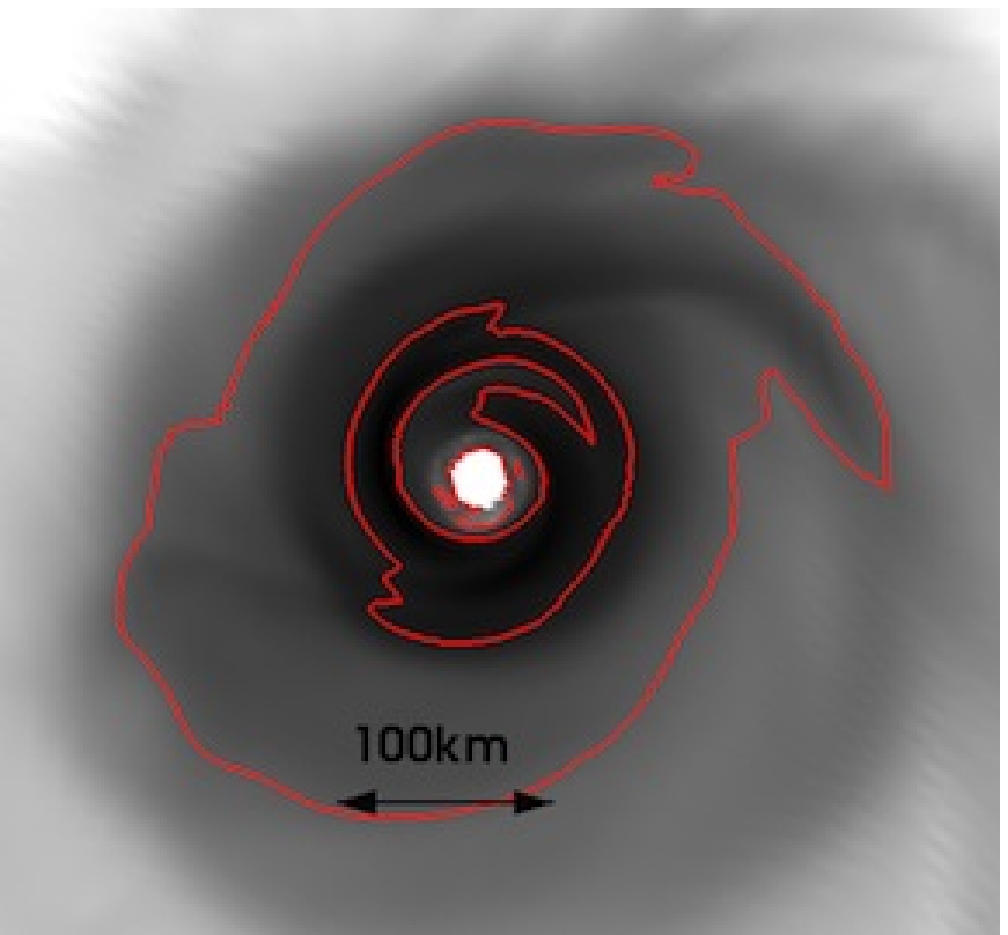}
\end{figure*}

This is no longer the case for the high spin configuration, $a_{\rm
BH}/M_{\rm BH}=0.9$. The disruption (Fig.~\ref{fig:Q7S9}, left) is
very similar to the previous case, with a long tidal tail forming
from the tidally disrupted star. However, now the disk can form much
closer to the black hole because the innermost circular orbit is at
a smaller radius. Also, the higher rotation speed of the hole prevents rapid
accretion of high angular momentum material. Thus the amount of nuclear
matter remaining outside of the black hole is now much larger:
about $30\%$ of the neutron star mass. Within approximately $5\,{\rm ms}$
of disruption, a large fraction of that material (about half)
forms an accretion disk (Fig.~\ref{fig:Q7S9}, center) with stable
density and angular momentum profiles.

The evolution of the resulting accretion disk is then similar
to what is expected for lower mass black holes. The disk expands
slightly, and heats up to $T\sim 5$ -- $6\,{\rm MeV}$, but does not evolve
much otherwise. The disk is thick ($H/R\sim 0.3$), and an order
of magnitude denser than for $a_{\rm BH}/M_{\rm BH}=0.7$. From
the accretion rate at the end of the simulation, we deduce an
expected lifetime of about $75\,{\rm ms}$, although this value
would certainly be modified by the inclusion of magnetic effects.
Another important feature of this high-spin case is that a few
percent of the total mass appears to be either unbound or on orbits
such that it will not interact with the accretion disk within
its expected lifetime. As the outermost part of the tidal tail is
poorly resolved in our simulations (the grid points are focused in
the region close to the black hole where the accretion disk forms),
predictions about the future of material ejected far away from the
black hole are not very reliable. The mere presence of such material,
however, indicates that massive ejecta might occur in BHNS mergers
if the mass and spin of the black holes are high.
By contrast, in previous
simulations for lower BH masses or spins, all of the material in
the tidal tail appeared unequivocally bound to the black hole.
%This point is discussed in Sec.~\ref{sec:ejecta}. 

\begin{table}
\caption{Parameters of the final configuration. $M_{\rm disk}$ is
defined as the mass available outside the black hole $5\,{\rm ms}$
after disruption. $a_{\rm BH}/M_{\rm BH}$ is the final dimensionless
spin of the black hole. $\rho_0^{\max}$ is the maximum baryon density
once the disk settles to an equilibrium configuration $20\,{\rm ms}$
after disruption. $T$ is the average temperature and $H$ is the
semi-thickness of the disk at that time. }
\label{tab:Res}
\begin{tabular}{|c|c|c|c|c|c|c|}
\hline
 Name & $\frac{M_{\rm disk}}{M_{\rm NS}^0} $ & $\frac{a_{\rm BH}}{M_{\rm BH}}$
& $\rho_0^{\max} (\frac{\rm g}{\rm cm^3})$ & $T ({\rm MeV})$ & $\frac{H}{r}$ &
$v_{\rm k} (\frac{\rm km}{\rm s})$\\
 \hline
Q7S5 & $\leq 0.004$ & 0.67 &   & Not Applicable &  & 86 \\
 \hline
Q7S7 & $0.06$ & 0.80 & $2\times 10^{10}$ & 2 & 0.3 & 63 \\
 \hline
Q7S9 & $0.28$& 0.92 & $3\times 10^{11}$ & 6 & 0.3 & 39\\
%\hline
%Q7S9c & $0$ & & \\
 \hline
Q5S5 & $0.06$  & 0.71 & $8\times 10^{11}$ & 2 & 0.3 & 106 \\
 \hline
\end{tabular}
\end{table}

\begin{figure}
\caption{Baryon mass available outside the black hole as a
function of time. The mass is normalized by the initial mass of
the neutron star.}
\label{fig:RestMass}
\includegraphics[width=8.3cm]{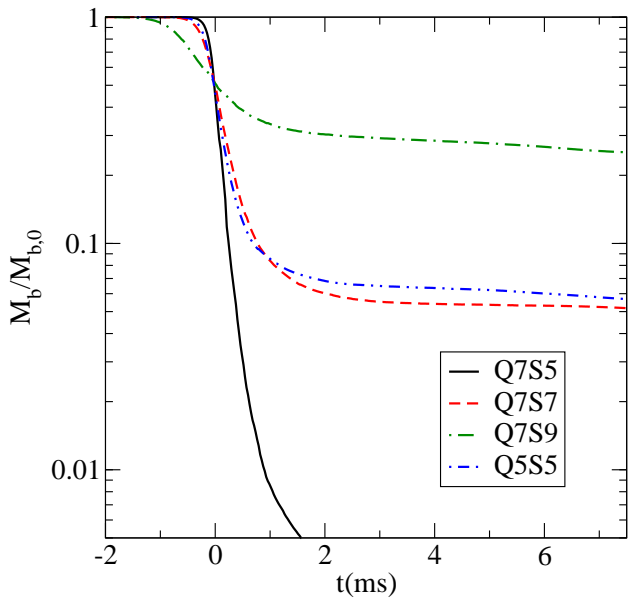}
\end{figure}

The large differences in the amount of matter available outside
the black hole as we modify the black hole spin can be seen in
Fig.~\ref{fig:RestMass}. The figure shows the difference between
a merger without disruption (Q7S5), with disruption followed by
the slow formation of a low density, short-lived disk (Q7S7),
and with disruption followed by the formation of a heavy, mostly
stable accretion disk (Q7S9). For $M_{\rm BH}/M_{\rm NS}=7$, we
should thus expect strong qualitative differences in the behavior
of BHNS mergers as we vary binary parameters such as the black
hole spin or the NS equation of state, with a significant
fraction of the parameter space now leading to direct merger
of the binary without the formation of an accretion disk. From
Fig.~\ref{fig:RestMass}, we can also easily evaluate the influence
of the black hole mass. The lower mass ratio simulation ($q=5$),
which has a black hole spin $a_{\rm BH}/M_{\rm BH}=0.5$, behaves
approximately as the $q=7$, $a_{\rm BH}/M_{\rm BH}=0.7$ case,
although the radius of the accretion disk around the less massive
black hole is naturally smaller, and the maximum baryon density
much higher. In Section~\ref{sec:discussion}, we discuss in more
detail how the qualitative behavior of BHNS mergers is affected
by both the mass and spin of the black hole, and how this relates
to their ability to power short gamma-ray bursts.

The results of the mergers are summarized in Table~\ref{tab:Res}. In
addition to the disk mass and properties discussed above,
Table~\ref{tab:Res} lists the final spin of the BH and the
velocity kick imparted to the hole. As expected, the relative
increase in the final black hole spin decreases as the initial
spin of the black hole increases. Starting from a spin $a_{\rm
BH}/M_{\rm BH}=0.5$ leads to a final BH spin of $0.67$, while an
original spin of $0.9$ only rises to $0.92$ after merger. For
low spins, the velocity kicks are close to binary black hole
predictions~\cite{2009CQGra..26i4023R,2011CQGra..28k4015Z}, but
as we go to higher spins, the magnitude of the kick decreases. This
could be expected, as most of the contribution to the kick in binary
black hole mergers comes from the momentum radiated as gravitational
waves just before merger, exactly the part of the signal that
is not emitted by BHNS systems when the star is tidally disrupted
before reaching the ISCO. The same effect was observed by Kyutoku
et al.~\cite{PhysRevD.84.064018} at lower mass ratios.

\subsection{Gravitational wave signal}
\label{sec:GW}

\begin{figure}
\caption{Gravitational wave signal for a mass ratio $q=7$,
when varying the black hole spin between $a_{\rm BH}/M_{\rm
BH}=0.5$ -- $0.9$. The time and phase of the waveforms are matched at
the peak of the signal.}
\label{fig:Q7Wave}
\includegraphics[width=8.3cm]{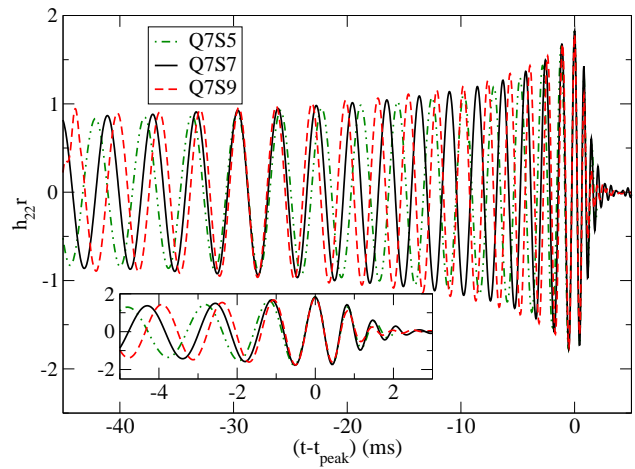}
\end{figure}

\begin{figure}
\caption{Power spectrum of the gravitational wave signal for a mass
ratio $q=7$, while varying the black hole spin between $a_{\rm
BH}/M_{\rm BH}=0.5$ -- $0.9$. The cutoff at low frequency is only an
effect of the finite length of the evolution. $h_{\rm eff}$ is the effective amplitude 
of the gravitational wave signal, as defined in Eq. (41) of~\cite{Shibata:2009cn}.}
\label{fig:Q7Spectrum}
\includegraphics[width=8.3cm]{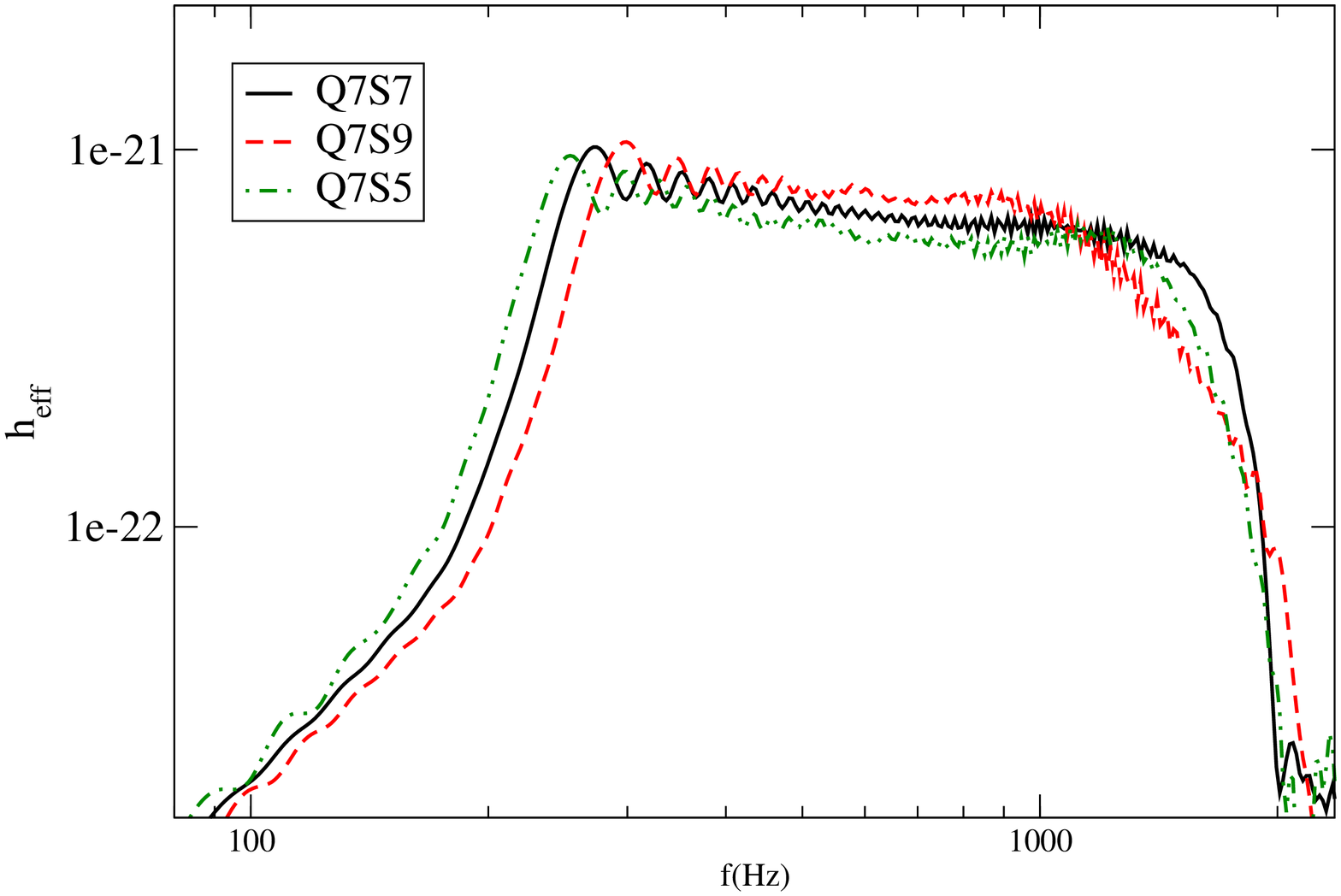}
\end{figure}

The gravitational wave signal emitted as two compact objects spiral
in and merge contains information on the orbital parameters of the
binary, the masses and spins of the compact objects, and, in the
presence of a neutron star, on the equation of state of matter
above nuclear density. In BHNS binaries, as in BH-BH systems,
these parameters leave a mark on the waveform as the binary spirals
in. But on top of the information from the orbital evolution
of the binary, BHNS systems also show strong variations in the
qualitative features of the gravitational wave signal at the time of
merger. This is visible on Fig.~\ref{fig:Q7Wave}, which shows the
gravitational strain measured for various binaries with $q=7$ but
different spins, and Fig.~\ref{fig:Q7Spectrum}, which shows the same
signals in the frequency domain. The low spin simulation Q7S5 has a
spectrum qualitatively similar to that of a BH-BH binary : the power
slowly decreases with increasing frequency as the binary spirals in,
then peaks at the time of merger ($\sim 1\,{\rm kHz}$), and finally falls
off exponentially as the remnant black hole rings down. At higher
spins ($a_{\rm BH}/M_{\rm BH} \geq 0.7$), the neutron star
disrupts and we no longer observe a peak in the gravitational wave
spectrum. The high-frequency signal now depends on the details of the
tidal disruption of the star. Lower spins lead to less disruption,
with most of the mass still falling rapidly into the black hole,
while higher spins lead to more significant disruption, and a more
homogeneous spread of the neutron star material around the black
hole. Accordingly, high spin spectra are cut off at lower frequency,
but fall off more smoothly. Similar qualitative differences are
found by Kyutoku et al.~\cite{PhysRevD.84.064018} and Etienne
et al.~\cite{Etienne:2008re}. In fact, Fig.~\ref{fig:Q7Spectrum}
is strikingly similar to Fig.~18 of~\cite{PhysRevD.84.064018},
which shows the spectra of the gravitational wave signal
for BHNS binaries with $q=3$ and spins $a_{\rm BH}/M_{\rm
BH}=-0.5,0,0.5$. Observationally, the effects of tidal disruption on
the waveform are slightly easier to measure in the case of a high
mass ratio, as the merger occurs at frequencies about a factor of
2 smaller than for $q=3$  ($f\sim 1$ -- $1.5\,{\rm kHz}$), but this still
remains a challenging frequency range for Advanced LIGO.

\begin{figure}
\caption{Gravitational wave signal for the low spin simulations
($a_{\rm BH}/M_{\rm BH}=0.5$) for mass ratios
$q=(5,7)$. The time and phase of the waveforms are matched at the
peak of the signal.}
\label{fig:Q5Wave}
\includegraphics[width=8.3cm]{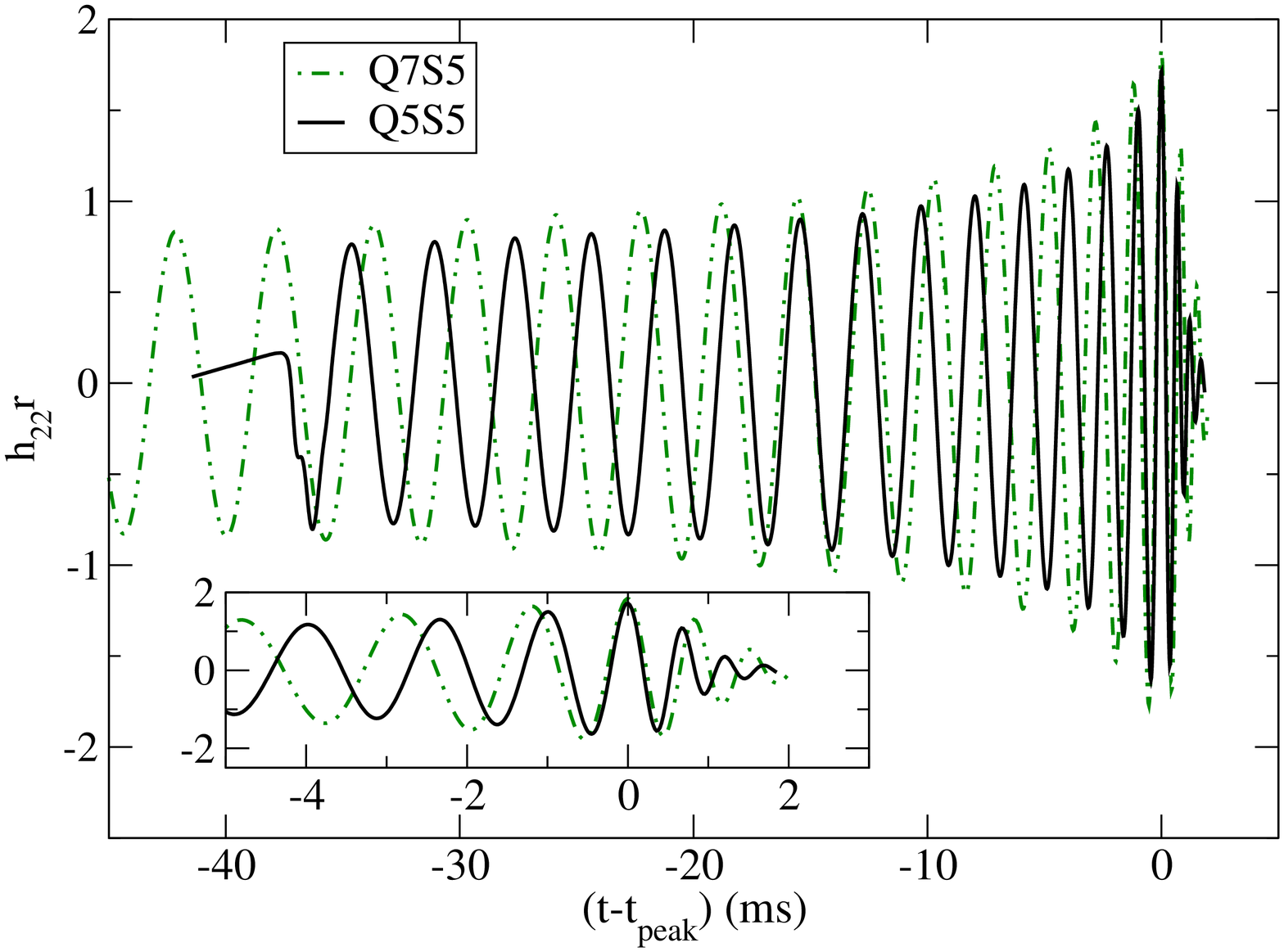}
\end{figure}

\begin{figure}
\caption{Power spectrum of the gravitational wave signal for the
low spin simulations ($a_{\rm BH}/M_{\rm BH}=0.5$)
for mass ratios $q=(5,7)$.}
\label{fig:Q5Spectrum}
\includegraphics[width=8.3cm]{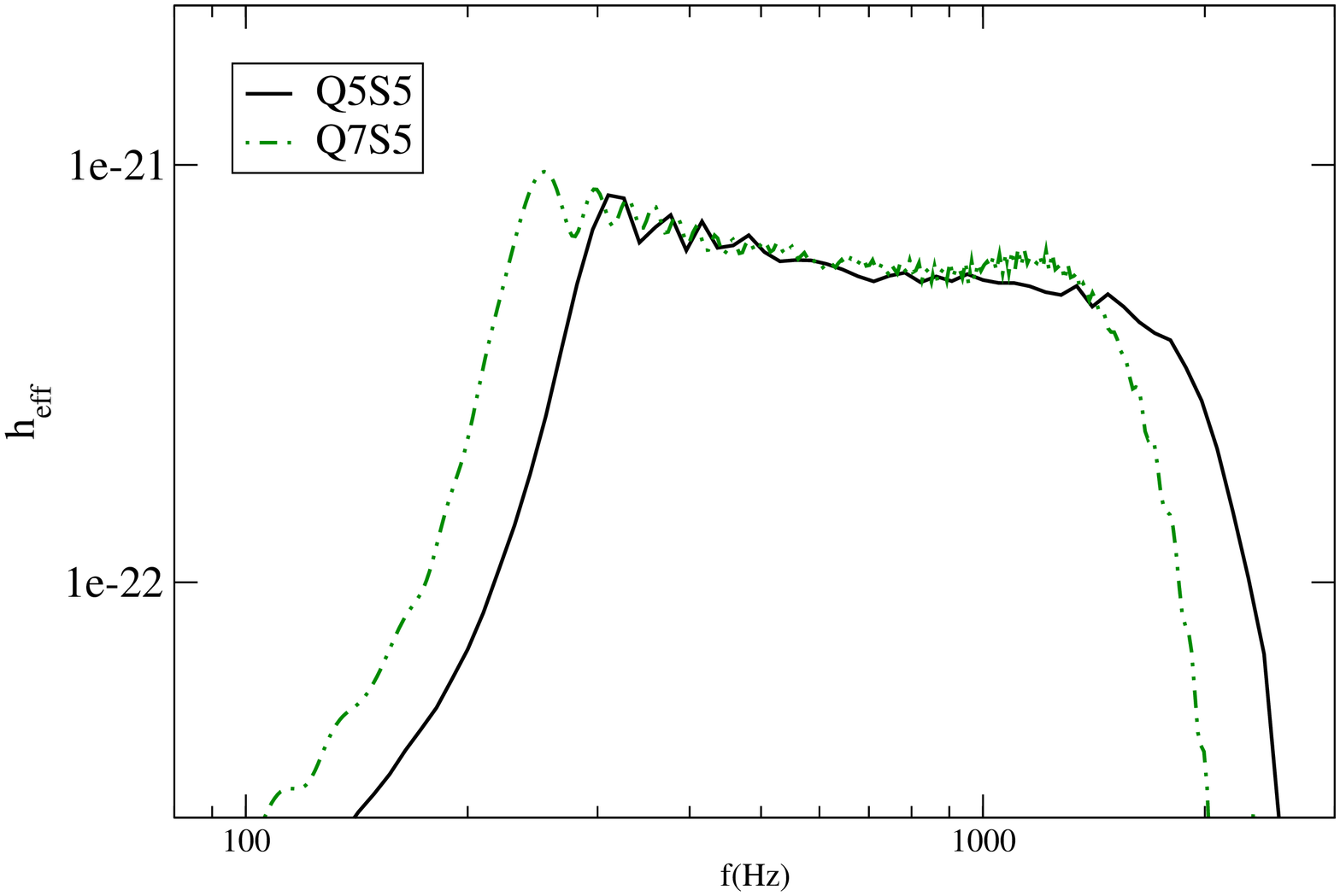}
\end{figure}

Figs.~\ref{fig:Q5Wave} and \ref{fig:Q5Spectrum} show the gravitational
wave signal of the two $a_{\rm BH}/M_{\rm BH}=0.5$ simulations in
the time and frequency domains. As expected from the above discussion,
the lower mass ratio has a behavior similar to the $q=7$, $a_{\rm
BH}/M_{\rm BH}=0.7$ case: both configurations have similar tidal
disruption history, with a few percent of the mass being ejected
in a tidal tail and the rest of the neutron star merging quickly
with the black hole. The signal in the lower mass case ($q=5$)
is simply at a slightly higher frequency.

It is worth noting that similar differences can
also be due to changes in the radius of the neutron
star~\cite{Shibata:2009cn,Duez:2010a,2010PhRvD..82d4049K,PhysRevD.84.064018}.
In order to extract information from that high-frequency signal, it
is thus important to already have good estimates of the parameters
of the binary to break that degeneracy. Given accurate
gravitational wave templates, this can be obtained from the signal
at lower frequency, during the orbital evolution.

\section{Limits on disk formation and Gamma-Ray Bursts}
\label{sec:discussion}

\begin{figure}
\caption{Simulations with similar mass remaining outside
the black hole for various BH masses and spins. All simulations
correspond to $M_{\rm NS}=1.4M_\odot$, $R_{\rm NS}=14.4\,{\rm km}$
(although a rescaling of $M_{\rm NS}$, $R_{\rm NS}$ and $M_{\rm BH}$
would lead to the same fraction of the stellar mass remaining at
late time). They thus correspond to optimistic predictions for the
formation of accretion disks.}
\label{fig:MRvsSpin}
\includegraphics[width=8.3cm]{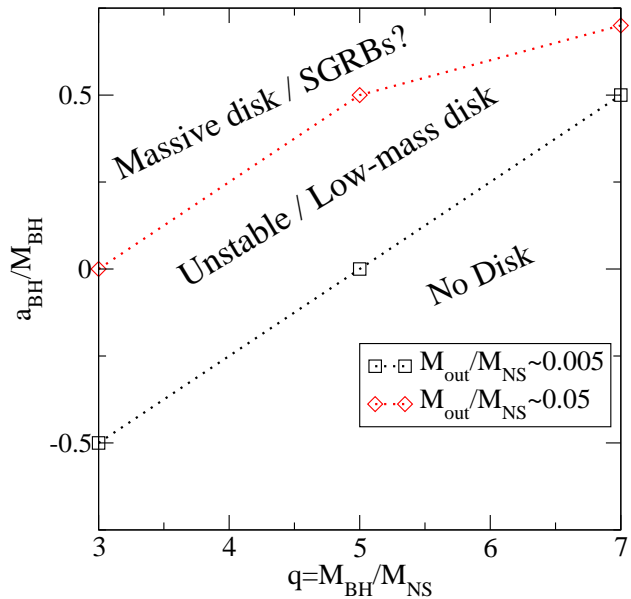}
\end{figure}

All the simulations presented in this paper use fairly large stars,
with $R_{\rm NS}=14.4\,{\rm km}$ for $M_{\rm NS}=1.4\,M_\odot$. They
correspond to particularly favorable cases for disk formation. They
thus allow us to deduce limits on the minimum spin required to form
significant accretion disks, at least if we assume that real neutron
stars have radii $R_{\rm NS}<14.4\,{\rm km}$ (as predicted by Hebeler
et al.~\cite{2010PhRvL.105p1102H}). Numerical simulations of BHNS
using nearly the same nuclear equation of state have already been
carried at lower mass ratios by Etienne et al.~\cite{Etienne:2008re},
as well as in our previous paper~\cite{2011PhRvD..83b4005F}. From
these results, we notice that the behavior observed here in the
cases $q=5$, $a_{\rm BH}/M_{\rm BH}=0.5$ and $q=7$, $a_{\rm BH}/M_{\rm
BH}=0.7$
is very similar to the results obtained for  $q=3$, $a_{\rm BH}/M_{\rm
BH}=0.0$~\cite{Etienne:2008re,2011PhRvD..83b4005F}: 
the amount of matter remaining outside the
black hole at late times is $M_{\rm out}/M_{\rm NS} \sim 0.05$ -- $0.06$.
However, these three
cases are not exactly equivalent, of course. The higher mass ratio
corresponds to a larger, lower density disk, which generally appears
less favorable to the generation of SGRBs. Nonetheless, all the cases
lead to a stable accretion disk of relatively low mass. Similarly,
we have three cases for which nearly the entire star is quickly
accreted onto the black hole, with a small but measurable
amount of material ejected in a tidal tail $M_{\rm out}/M_{\rm
NS} \sim 0.005$ -- $0.008$: the  $q=7$, $a_{\rm BH}/M_{\rm BH}=0.5$
case presented here, as well as the $q=5$, $a_{\rm BH}/M_{\rm
BH}=0.0$ and  $q=3$, $a_{\rm BH}/M_{\rm BH}=-0.5$ cases of Etienne et
al.~\cite{Etienne:2008re}. We thus have two curves in the space of
black hole masses and spins giving us cases where the amount of
matter available at late time for a $R_{\rm NS}=14.4\,{\rm km}$,
$M_{\rm NS}=1.4M_\odot$ star is $\sim 0.01M_\odot$ and $\sim
0.08M_\odot$ respectively (see Fig.~\ref{fig:MRvsSpin}). Below the
lowest curve of Fig.~\ref{fig:MRvsSpin}, no significant disruption
of the neutron star occurs before merger. Tidal effects are likely
to affect the gravitational wave signal slightly, but the formation
of an accretion disk is not possible. Above the highest curve,
massive disks are formed for this particular stellar radius. However,
all that can be said is that massive disks are possible: the more
compact the neutron star, the more these curves would be displaced
toward higher black hole spins. Finally, in between the two curves,
we form a low mass disk of fairly low density or just a tidal tail
containing a few percent of the initial neutron star mass and
slowly falling back onto the central black hole.

In order for a BHNS binary to generate a short gamma-ray
burst, a necessary condition would thus be to lie in the
upper region of Fig.~\ref{fig:MRvsSpin}. This is not, however,
sufficient. More compact stars make it more difficult to form
accretion disks(see e.g.~\cite{2011ApJ...727...95P} for a model based on earlier, 
lower mass ratio simulations), and the same is true if the spin of the black
hole and the orbital angular momentum are not aligned, as
shown in Foucart et al.~\cite{2011PhRvD..83b4005F} for $q=3$
mass ratios. In~\cite{2011PhRvD..83b4005F}, the effect of
spin misalignment was fairly limited, but we were considering
configurations for which even a nonspinning black hole led to
the formation of a significant accretion disk. Nevertheless, the
final disk mass appeared to depend on the aligned component of
the black hole spin. There is no guarantee that this will still
be true for higher mass ratio, or for nearly extremal spins;
further numerical studies in a general relativistic framework
will be necessary to determine the effect of spin misalignment
in BHNS binaries for $q\sim 7$ -- 10. But if the same pattern
holds for $q=7$ and $q=3$, misalignment angles $\sim 40^\circ$
would be enough to have a BHNS with spin $a_{\rm BH}/M_{\rm
BH}=0.9$ behave as an aligned configuration with $a_{\rm BH}/M_{\rm
BH}=0.7$. As misalignment angles of $\sim 40^\circ$ could be fairly
common~\cite{2008ApJ...682..474B}, this would affect the probability
that BHNS mergers form massive accretion disks.

We should also note that BHNS binaries which do not form an accretion 
disk could still emit a detectable electromagnetic signal before the disruption 
of the neutron star, for example in the form of resonant shattering precursor 
flares~\cite{2012PhRvL.108a1102T}.

\section{Conclusion}
\label{sec:conclusion}

Population synthesis models indicate that existing simulations of
BHNS mergers in general relativity have probably studied black holes with
masses below the most common astrophysical values. Binaries with
lower mass black holes could occur, and their study has already
offered valuable insights into the influence of various binary
parameters such as BH spin, NS equation of state and mass ratio on
the emitted gravitational wave signal and the disk formation
process. However, studies of the behavior of BHNS binaries for more
massive black holes are necessary if we want to obtain more accurate
predictions for what is likely to be a significant portion of the
available BHNS parameter space.

In this paper, we present the first simulations of BHNS binaries
with $M_{\rm BH} = 10 M_\odot$. We show that, as opposed to what
was found for lower mass black holes, the tidal disruption of the
neutron star and the formation of an accretion disk are no longer
the norm. We only observe significant accretion disks for high
black hole spins, $a_{\rm BH}/M_{\rm BH} \geq 0.7$, and this
result is obtained using a fairly large neutron star ($R_{\rm NS}
\sim 14.4\, {\rm km}$), thus providing a likely lower bound on the
spin required to form accretion disks in such systems.

As spin-orbit misalignment also inhibits disk
formation~\cite{2011PhRvD..83b4005F}, it appears that the formation
of massive accretion disks will be a common result of BHNS mergers
only if large black hole spins aligned with the orbital angular
momentum are frequent, or if black holes in BHNS binaries are less
massive than expected.
This is of course of particular importance when determining
whether BHNS mergers can be the progenitors of short gamma-ray
bursts. BHNS systems with $M_{\rm BH} \sim 10 M_\odot$ and $a_{\rm
BH}/M_{\rm BH} \leq 0.7$ appear to be unlikely candidates to power
SGRBs. In Sec.~\ref{sec:discussion}, we used existing simulations of
BHNS binaries for mass ratios $q=3$ -- 7 to obtain approximate lower
bounds for the magnitude of the black hole spin required to obtain
massive disks.

BHNS systems with $M_{\rm BH} \geq 7 M_{\rm NS}$ show another
difference with respect to previously studied binaries: for high
black hole spins $a_{\rm BH}/M_{\rm BH} \sim 0.9$, a few percent
of the NS is either unbound or weakly bound. This could lead to the
ejection of neutron-rich material in the neighborhood of the binary.
However, study of this effect will require the use of numerical
methods that resolve the evolution of the tidal tail better than
our current code.

We showed that we can extract gravitational waves with a cumulative
phase error $\Delta \phi < 0.2\, {\rm rad}$ over about five orbits,
without applying any time or phase shift to the signal. When matching
the time and phase of the signal at its peak, our accuracy is similar
to that reported by Kyutoku et al.~\cite{PhysRevD.84.064018}
for lower mass black holes. The gravitational wave signal shows
that as the spin of the black hole decreases,
the spectrum goes through three regimes. At high spins, the spectrum is
cut off at low frequency as the star is disrupted and the remaining
material forms a nearly axisymmetric accretion disk.
For intermediate spins, the spectrum is mostly
flat, extending to frequencies $\sim 2\,{\rm kHz}$
at which disruption leads to the formation of a thin
tidal tail.
At low spins, the spectrum is very similar to the signal from BH-BH binaries,
where tidal disruption does not occur.

The behavior of BHNS binaries for higher mass ratios ($q>7$) or
smaller stellar radii should be even more similar to BH-BH binaries,
with small corrections for the tidal distortion of the star. Whether
any such binary can form an accretion disk will depend on how mergers
occur for quasi-extremal spins, and how frequent high-spin black
holes are in BHNS binaries. For high-spin systems, the relative
alignment of the black hole spin and the orbital angular momentum
is also likely to play an important role. In an earlier study of
BHNS binaries with misaligned spins~\cite{2011PhRvD..83b4005F},
we found that the qualitative effect of misalignment was relatively
modest, but so were the effects of the spin itself, as even a
nonspinning BH led to the formation of an accretion disk. At high mass
ratios, it is likely that tidal tail and disk formation will only be
possible for small spin-orbit misalignments, as is also predicted
by Newtonian simulations~\cite{2008ApJ...680.1326R}. Additional
simulations will, however, be necessary to obtain more accurate
predictions for such systems.

\acknowledgments
We thank Dan Hemberger, Jeff Kaplan, Geoffrey Lovelace, 
Curran Muhlberger and Harald Pfeiffer for useful discussions and suggestions. 
This work was supported in part by a grant from the Sherman Fairchild
Foundation; by NSF Grants Nos.\ PHY-0969111 and PHY-1005426 and NASA Grant No.
NNX09AF96G at Cornell; and by NSF Grants Nos.\ PHY-0601459, PHY-1068881 and
PHY-1005655 and NASA Grant No.\ NNX09AF97G at Caltech.
M.D. acknowledges support through NASA Grant No.\ NNX11AC37G
and NSF Grant PHY-1068243.  This research
was supported in part by the NSF through
TeraGrid~\cite{teragrid} resources provided by
NCSA's Lonestar cluster under Grant No. TG-PHY990007N.
Computations were also performed on the Caltech compute cluster
``Zwicky'', funded
by NSF MRI award No. PHY-0960291, and on the GPC
supercomputer at the SciNet HPC Consortium.  SciNet is funded by: the
Canada Foundation for Innovation under the auspices of Compute Canada;
the Government of Ontario; Ontario Research Fund - Research Excellence;
and the University of Toronto.

\bibliography{References/References}

\end{document}